# Resistively detected electron spin resonance and *g*-factor in few-layer exfoliated MoS$_2$ devices

Chithra H. Sharma[1,2*], Appanna Parvangada[2], Lars Tiemann[2], Kai Rossnagel[1,3], Jens Martin[4], Robert H. Blick[2,5]

[1]*Institut für Experimentelle und Angewandte Physik, Christian-Albrechts-Universität zu Kiel, 24098 Kiel, Germany*
[2]*Center for Hybrid Nanostructures (CHyN), Universität Hamburg, Luruper Chaussee 149, 22761 Hamburg, Germany*
[3]*Ruprecht Haensel Laboratory, Deutsches Elektronen-Synchrotron DESY, 22607 Hamburg, Germany*
[4]*Leibniz Institut für Kristallzüchtung, 12489 Berlin, Germany*
[5]*Material Science and Engineering, University of Wisconsin-Madison, University Ave. 1550, Madison, 53706, Wisconsin, USA*

*\*Chithra.Sharma@physik.uni-hamburg.de*

**Abstract**

MoS$_2$ has recently emerged as a promising material for enabling quantum devices and spintronic applications. In this context, an improved physical understanding of the *g*-factor of MoS$_2$ depending on device geometry is of great importance. Resistively detected electron spin resonance (RD-ESR) could be employed to and the determine the *g*-factor in micron-scale devices However, its application and RD-ESR studies have been limited by Schottky or high-resistance contacts to MoS$_2$. Here, we exploit naturally *n*-doped few-layer MoS$_2$ devices with ohmic tin (Sn) contacts that allow the electrical study of spin phenomena. Resonant excitation of electron spins and resistive detection is a possible path to exploit the spin effects in MoS$_2$ devices. Using RD-ESR, we determine the *g*-factor of few-layer MoS$_2$ to be ~1.92 and observe that the *g*-factor value is independent of the charge carrier density within the limits of our measurements.

## 1. Introduction

Semiconducting transition-metal dichalcogenides (TMDCs), a family of van der Waals materials have gained attention in spintronics and quantum computing technology due to their electronic and spin-based properties. [1–3] Recent progress made in device engineering and wafer-scale device integration has positioned 2D-TMDCs, especially MoS$_2$, on the industrial roadmap making them a promising material platform for future quantum devices and technologies. [4,5] MoS$_2$ is of prime interest due to its gate tunability, chemical, mechanical and electrical stability, and natural abundance. While there are many demonstrations of microelectronic circuits, MoS$_2$ has also been identified as a platform for optoelectronic and spintronic technologies. [1,2] MoS$_2$ has shown strong spin-valley interactions in monolayer and twisted bilayer MoS$_2$ making them also a suitable candidate for valleytronics. [6–8] It has also exhibited various quantum transport phenomena such as weak localization, Shubnikov de Haas oscillations, conductance quantization on gated quantum point contacts etc. [9,10]. Recent demonstrations of electrostatically defined single and double quantum dots along with the proposals for spin, valley and Kramer's qubits place MoS$_2$ as a future platform for gated quantum dot-based qubits. [11,12]

Landé *g*-factor is arguably the most important parameter one needs to get a handle on as far as spin-based devices are concerned. For monolayer MoS$_2$, theory predicts an electron *g*-factor in the range of 1.98 to 2.2 for the conduction band. [12] There are reports of *g*-factor measurements on bulk impurity-doped MoS$_2$ samples utilizing conventional absorption electron spin resonance (ESR) spectroscopy. [13,14] Using conventional ESR spectroscopy on multilayer chemical vapor deposited (CVD) MoS$_2$ devices with ionic gel gating, temperature-dependent *g*-factors for various vacancies and defects in the range of 2.045 to 2.055 have been determined. [15] Alternatively, time-resolved Kerr rotation spectroscopy on CVD-grown *n*-type $MoS_2$ revealed a *g*-factor in the range of 1.902 to 1.975. [16] Additionally, electronic transport in 1D-channels of split-gated $MoS_2$ devices was investigated and a *g*-factor of 2.16 was extracted from bias spectroscopy. [17] However, a strongly confined geometry can significantly affect the *g*-factor, so the value cannot be assumed to be entirely an intrinsic property of the material. [18] As a result, there is no clear consensus on the actual *g*-factor value of $MoS_2$, specifically for device applications. Conventional measurements are performed on bulk samples or on a large number of flakes, and the techniques used are not suitable for probing low-dimensional systems. They cannot be extended to quantum devices such as gated quantum dots and quantum point contacts, which are submicron structures realized in monolayer and few-layer devices, nor can they be used to control spins in such devices.

Here, we adopt resistively detected ESR (RD-ESR) measurements which allow *in operando* studies of real device geometries. In this technique, the population of the Zeeman split states in an applied magnetic field *B* is manipulated by applied microwave radiation of frequency $\nu$ causing resonant changes in the resistance when the resonance condition is met. [6,19] It can be used to drive spin rotations and, as a tool for extracting the *g*-factor. In fact, RD-ESR has been used to determine the *g*-factor in monolayer graphene and to measure the effects of intrinsic spin-orbit coupling and pseudospin in graphene devices. [19–21] In graphene/$MoS_2$ heterostructures, the modulation of the *g*-factor of graphene in proximity to $MoS_2$ has been reported. [6] RD-ESR has also been used to probe confined electron systems to understand the modulation of the *g*-factor as a result of the confinement and electric field tunability aiding in the design of quantum dot qubits. [17,18] Recently, RD-ESR has also emerged as a tool to drive spin rotations, making it attractive for the control of spin qubits. [22]

Measurement of electron *g*-factor in $MoS_2$ using RD-ESR is highly desirable, yet. RD-ESR experiments require high sensitivity in the measurement of the electrical current/voltage to detect the subtle resistance changes. This makes devices with low contact resistances a prerequisite. Especially for low-temperature experiments, large contact resistances and Schottky barriers have prevented direct *g*-factor measurements in $MoS_2$ by RD-ESR. Most common metals used in device fabrication such as gold, are known to form large Schottky barriers on $MoS_2$ due to the work function mismatch, inefficient orbital overlap or Fermi-level pinning due to the material damage at the interface during the metal deposition process. [23,24] Recently, metals with lower melting points and lower work functions, including indium (In) and tin (Sn), have been shown to form ohmic contact with $MoS_2$. [23,25–27] Specifically, Sn contacts deposited at low temperatures have shown ohmic contact resistances of only ~1-3 kΩ µm. [26,27] In this work, we use thermally evaporated Sn deposited on liquid $N_2$-cooled samples. [26] Our few-layer $MoS_2$ devices show excellent device performance, as demonstrated and discussed below. We perform magneto-transport measurements and RD-ESR spectroscopy [19] to extract the electronic $g$-factor in the conduction band of a few-layer $MoS_2$ field effect transistor device at ~1.6 K. We were able to measure the $g$-factor for a range of charge carrier densities by resistive detection. Our measurements also demonstrate the potential use of RD-ESR for spin manipulation of $MoS_2$-based devices.

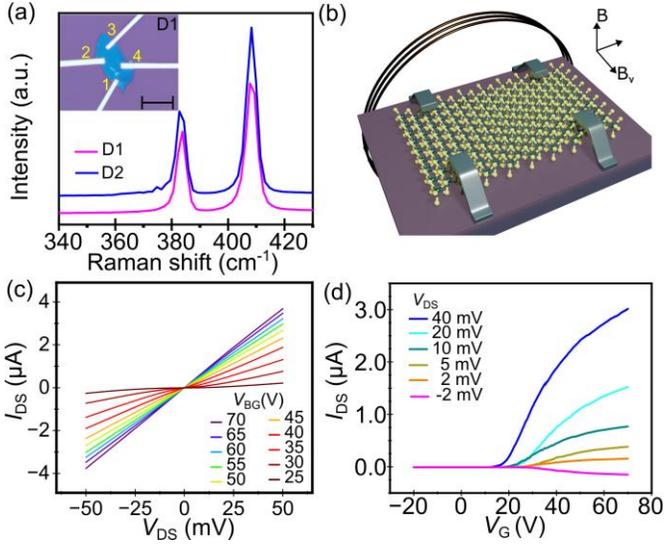

Fig. 1: (a) Raman spectra of the $MoS_2$ flakes showing the $E^1_{2g}$ and $A_{1g}$ peaks. Inset: optical image of the device D1 with Sn contacts and the contacts labeled from 1 to 4. Scale bar is 20 μm. (b) Schematic of the experimental setup showing the device with the loop antenna (not to scale). (c) Two-probe *I-V* curves at ~1.6 K for different back-gate voltages ($V_{BG}$) for D1. (d) Current as a function of gate voltage of the device D1 for different bias voltages ($V_{DS}$) at ~1.6 K.

## 2. Results and Discussion

### 2.1 Device fabrication and basic characterisation

We have investigated two devices labelled as D1 and D2. The devices are fabricated by manually exfoliating few-layer $MoS_2$ flakes and transferring them to a $p$-doped Si wafer with 300 nm $SiO_2$ and prefabricated Ti/Au bonding pads, using a PDMS-assisted dry transfer technique. [28] We used commercially available natural $MoS_2$ bulk material (SPI supplies). The flakes were identified by optical contrast to be 8-12 layers thick. [29] The ohmic contacts to $MoS_2$ were realized by e-beam lithography followed by Sn evaporation and lift-off. During the Sn evaporation, the sample stage was cooled with liquid $N_2$ for a smooth metal film formation and to avoid kinetic damage of $MoS_2$. [23,26] Figure 1(a) shows the Raman spectra of the flakes in the device D1 (magenta) and D2 (blue), where the positions of the $E^1_{2g}$ and $A_{1g}$ peaks, obtained by Gaussian fits to the peaks, can be read as 383.49 $cm^{-1}$ and 408.27 $cm^{-1}$, respectively. The Raman spectra were recorded after the Sn deposition, indicating that the quality of the material was not compromised during the fabrication process. The optical image of device D1 is shown in the inset with the Sn contacts labelled 1-4. After wire bonding, the devices were mounted on a vacuum probe (~$10^{-5}$ mbar) with minimal exposure to the atmosphere (<30 min) to avoid oxidation of the metal-semiconductor interface. Electrical transport measurements were performed at a temperature of ~1.6 K in a cryostat equipped with a variable temperature insert and a superconducting solenoid magnet. The magnetic field, $B$, is applied perpendicular to the sample. A loop antenna of ~5mm diameter is placed adjacent to the device for microwave irradiation with the magnetic field vector of the photon, $B_v$, pointing perpendicular to the applied external magnetic field. A schematic (not to scale) of the device with a loop antenna is shown in Fig. 1(b) with the magnetic field direction marked. We first discuss the results for the D1 device. Figure 1(c) shows the current-voltage ($I$-$V$) characteristics of the device for different gate voltages, and Fig. 1(d) shows the two-probe (contacts 2 and 4) transfer characteristics at 1.6 K for different bias voltages. A linear behaviour of the $I$-$V$ characteristics can be observed for bias voltages in the mV range and for gate voltages of 45 V and above, where a saturation behaviour of the current can also be seen in the transfer characteristics. In the presence of Schottky contacts this saturation behaviour is not seen due to the influence of the gate on the contacts.

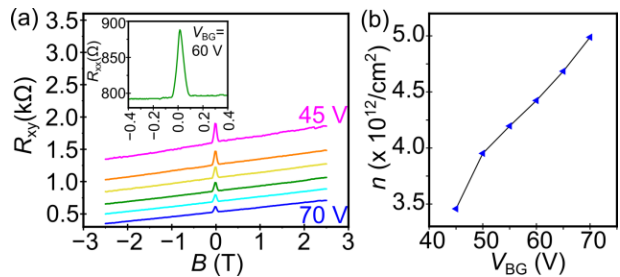

Fig. 2: (a) Magnetotransport measurement in Hall configuration for different gate voltages. The inset shows the weak-localization behavior in longitudinal mode at $V_{BG}$ = 60.0 V. (b) Extracted carrier densities as a function of the gate voltage.

We performed four-probe magnetotransport measurements at 1.6 K in the longitudinal ($I_{14}$, $V_{23}$) and transverse/Hall ($I_{24}$, $V_{13}$) configurations. The Hall measurements for different gate voltages for D1 are shown in Fig. 2(a). Weak localization behaviour, was observed in $R_{xx}$, as shown in the inset. Due to the asymmetry of the contacts, a small longitudinal component leads to an offset and a superimposed weak localization peak in the Hall effect measurement. Such a narrow weak localization peak indicating good device quality is typically seen only in encapsulated devices. [30–32] We extract carrier densities from the slope of the Hall data in the range of ~3.5-5 × $10^{12}$ $cm^{-2}$ for gate voltages of 45-70 V for D1 as shown in Fig. 2(b). Only the gate voltage range where linear $I$-$V$ is observed is considered here. The values are comparable to previously observed results in Sn-contacted $MoS_2$ devices. [26] The electron mobility was extracted using the formula $\mu = \frac{\sigma}{ne}$, where $n$ is the calculated carrier density, $e$ is the electronic charge, and $\sigma$ is the conductivity. We used $\sigma = \frac{l}{w}\frac{1}{R_{xx}(B=0)}$, where $l = 33$ μm is the distance between contacts 1 and 3 and $w = 15$ μm is the distance between contacts 2 and 4. For this device, we calculate maximum mobility of ~2900 cm$^2$/Vs at $V_{BG}$ = 70.0 V at 1.6 K, which is much higher than values reported in the literature for unencapsulated devices [30,33] and comparable to many encapsulated devices. [8,31,34] We note that our measurement geometry deviates from the standard Hall bar and van der Pauw geometry. Therefore, the estimated mobility should only be taken as an upper limit, while the true mobility may be lower by a factor of 4. We did not increase the gate voltage beyond 70 V to avoid gate leakage and dielectric breakdown in this device.

**2.2 RD-ESR measurements**

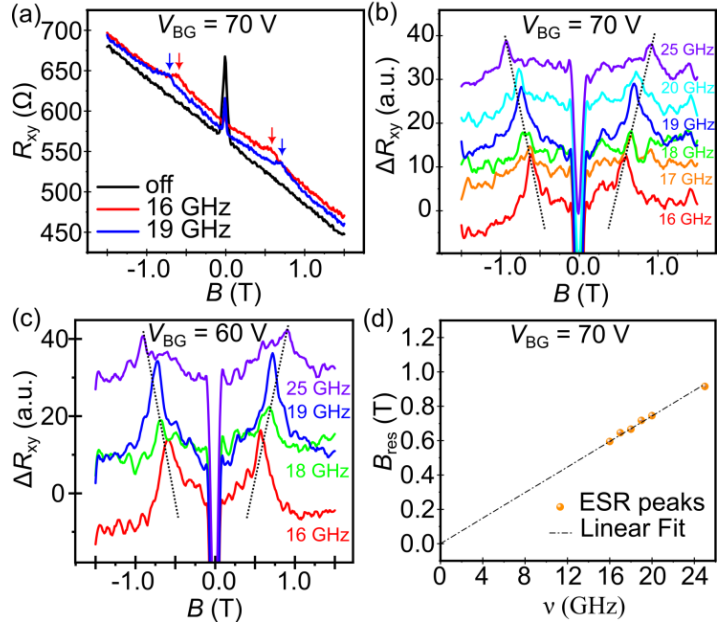

Fig. 3: (a) Magnetotransport data in the Hall configuration for microwave excitation off (black), 16.0 GHz (magenta), and 19.0 GHz (red) for D1 at $V_{BG}$ = 70.0 V. Background subtracted ESR data for different microwave excitations, offset for clarity, for D1 at gate voltages (b) $V_{BG}$ = 70.0 V, (c) $V_{BG}$ = 60.0 V. The dotted lines are guides to the eye following the ESR peaks (d) Extracted ESR peak positions for different excitation frequencies from (b) and the linear fit (dotted line) to extract the g-factor.

For the RD-ESR spectroscopy, we applied a microwave signal in the range of 15-26 GHz to the loop antenna as shown in the schematics in Fig. 1(b). A small change in the device resistance is observed due to resonant absorption of the spins in the magnetic field when $h\nu = g\mu_B B_{res}$, where $\nu$ is the applied frequency, $B_{res}$ is the external magnetic field at resonance, $h$ and $\mu_B$ are the Planck constant and Bohr magneton, respectively. Magnetotransport data in the Hall configuration with microwave excitations of 16 GHz (red) and 19 GHz (blue) are shown in Fig. 3(a) for device D1 at a gate voltage of 70.0 V. The curve in black shows the background where the microwave excitation is off [same as the blue curve in Fig. 2(a)]; i.e., $R_{xy}(\text{off})$. Clear ESR signals on the red and blue curves are indicated by the arrows of the respective colours. We chose the Hall configuration for the RD-ESR measurements because the linear background facilitates clear observation of the ESR peaks and maintains consistency across devices. A change in the total device resistance is also observed during microwave excitation due to the heating effect. For further analysis, background subtraction [i.e., $R_{xy}(\nu) - R_{xy}(\text{off})$] was performed on the RD-ESR data, where Figs. 3(b) and 3(c) show results for different microwave excitations on device D1 for $V_{BG}$ = 70.0 V and $V_{BG}$ = 60.0 V, respectively. The curves are offset for visibility and the dotted lines following the peaks are guides to the eye. The curves are smoothed (Savitzky-Golay) and Lorentzian fits to the resonance peaks were used to extract the $B_{res}$ for different frequencies. The extracted $B_{res}$ from Fig. 3(b), plotted against the microwave excitation frequencies, are shown in Fig. 3(d). The dotted line in Fig. 3(d) is a linear fit passing through the origin. From the slope, we extract a g-factor of 1.92±0.03 for D1 at a gate voltage of 70.0 V. Similarly, we estimate a g-factor of 1.91±0.01 for $V_{BG}$ = 60.0 V.

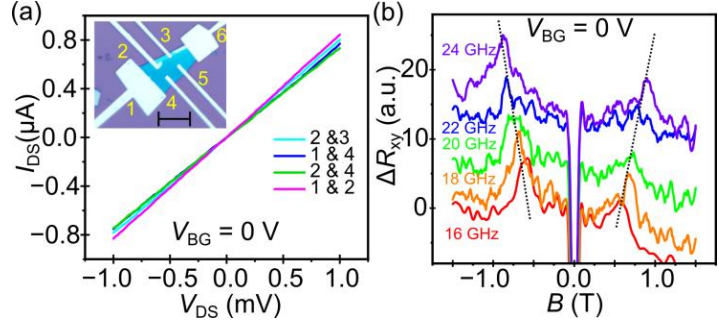

Fig. 4: (a) Two-probe I-V curves at ~1.6 K for different contact pairs at $V_{BG}$= 0 V for D2. Optical image of D2 in the inset to (d); scale bar is 10 µm. (b) Background subtracted data for different microwave excitations for D2 at $V_{BG}$ = 0.0 V.

We also performed the measurements on a second device, D2. Fig. 4(a) shows the two-probe I-V curves of D2 at 1.6 K for different pairs of contacts at $V_{BG}$ = 0 V. The inset shows the optical image of the device with contacts labelled from 1 to 6. Contacts 5 and 6 were not functional. Interestingly, D2 showed a linear I-V response even at $V_{BG}$ = 0 V, as shown in Fig. 4(a). Although there were no deliberate differences in the processing of the different devices, we assume that the local differences in the bulk material or some differences in the contact formation could be the probable reasons for this behaviour. Unfortunately, the device showed an onset of gate leakage above $V_{BG} \approx$ 2V, and hence the gate was kept at $V_{BG}$ = 0.0 V for further measurements to avoid any dielectric breakdown. We performed magnetotransport and RD-ESR measurements in the transverse/Hall ($I_{13}$, $V_{24}$) configuration at $V_{BG}$ = 0.0 V for D2. In this case, a carrier density of 2.9 x $10^{12}$ cm$^{-2}$ was measured (data not shown). Figure 4(b) shows the background subtracted ESR data and we estimate a g-factor of 1.92±0.02 at $V_{BG}$ = 0.0 V, as described above. The curves are offset for visibility and the dotted lines following the peaks are guides to the eye.

Our measurements on two devices cover a range of carrier densities. However, we observed that the value of the g-factor does not seem to have a significant dependence on the range of carrier densities studied, as summarized in Tab. 1. The variation of the charge carrier densities across the devices is consistent with the gate capacitance of the dielectric, i.e., $n = \frac{C_{ox}}{e} V_{BG}$, where $C_{ox}$ is the gate capacitance ($C_{ox} = \frac{\epsilon_r \epsilon_0}{d}$) and $e$ is the electronic charge, $\epsilon_r = 3.9$ is the dielectric constant for SiO$_2$ and $d = 300\ nm$ is the thickness of the dielectric. Hence, we can assume that the two devices have the same doping.

| Device | $V_{BG}$ (V) | Carrier density (cm$^{-2}$) | g-factor |
|---|---|---|---|
| D1 | 70.0 | (4.988±0.001) x $10^{12}$ | 1.92±0.03 |
| D1 | 60.0 | (4.423±0.001) x $10^{12}$ | 1.91±0.01 |
| D2 | 0.0 | (2.902±0.009) x $10^{12}$ | 1.92±0.02 |

Tab. 1: Summary of measured g-factors.

## 3. Conclusions

In summary, our experiments on specially tailored MoS$_2$ devices with ohmic contacts demonstrate that RD-ESR can be used to resistively detect and control spin effects in the material, enabling advanced studies for spintronics and qubit applications. In transport measurements the nature of the source-drain contacts plays an important role in unravelling the intrinsic properties of the materials. The ohmic contact provided by Sn and the excellent device quality allowed us to perform the RD-ESR measurements. We have measured the effective g-factor in few-layer MoS$_2$ to be ~1.92. We find a g-factor that is significantly lower than theoretical predictions as well as most other experimental results using other gating techniques. Similar measurements on graphene have also shown different g-factors than those measured by RD-ESR. [19,35] Variation in the measurement environment and the material could be a reason for this and that needs to be further investigated. However, most MoS$_2$ devices used for electron confinement or spintronics rely on Si/SiO$_2$ substrates and conditions similar to ours. We believe that our results will be valuable for the development of spin-based devices on MoS$_2$.

Considering the significant spin-orbit coupling (SOC) [8] in the conduction band (although much lower than in the valence band), the *g*-factor is surprisingly close to that of graphene (~1.95) and the free-electron *g*-factor (~2.0023). [19] While SOC is a determining factor, other factors such as spin-valley coupling, hyperfine coupling, exchange interactions, defects and impurities would also influence the spin properties. We expect that our measurements will pave the way for further studies to shed light on these as well as *g*-factor anisotropy, substrate dependence, the effect of doping, strain, SOC gap etc., which need to be explored in more detail in MoS$_2$.

## Acknowledgements

We thank Christian Schäfer (IKZ Berlin) for tin (Sn) deposition. We thank the Excellence Cluster CUI-AIM for support (EXC 2056).

## Author Declarations

The authors have no conflicts to disclose.

## Data availability statement

The data that support the findings of this study are available from the corresponding author upon reasonable request.